\documentclass[reprint,amsmath,amssymb,aps,prr,fleqn,showpacs]{revtex4-1}
\usepackage{graphicx}
\usepackage{enumerate}
\usepackage{dcolumn}
\usepackage{bm}
\begin{document}
\title{Three-dimensional structure of a string-fluid complex plasma}
\author{M.$\,$Y.$\,$Pustylnik$^1$}
\email{mikhail.pustylnik@dlr.de}
\author{B.$\,$Klumov$^{1,2,3}$}
\author{M.$\,$Rubin-Zuzic$^1$}
\author{A.$\,$M.$\,$Lipaev$^{2,3}$}
\author{V.$\,$Nosenko$^1$}
\author{D.$\,$Erdle$^1$}
\author{A.$\,$D.$\,$Usachev$^2$}
\author{A.$\,$V.$\,$Zobnin$^2$}
\author{V.$\,$I.$\,$Molotkov$^2$}
\thanks{deceased}
\author{G. Joyce}
\thanks{deceased}
\author{H.$\,$M.$\,$Thomas$^1$}
\author{M.$\,$H.$\,$Thoma$^4$}
\author{O.$\,$F.$\,$Petrov$^{2,3}$}
\author{V.$\,$E.$\,$Fortov$^2$}
\author{O.$\,$Kononenko$^5$}
\affiliation{$^1$Institut f\"ur Materialphysik im Weltraum, Deutsches Zentrum f\"{u}r Luft- und Raumfahrt (DLR), M\"{u}nchener Stra\ss e 20, 82234 We\ss ling, Germany}
\affiliation{$^2$Institute for High Temperatures, Russian Academy of Sciences, Izhorskaya 13/19, 125412 Moscow, Russia}
\affiliation{$^3$Moscow Institute of Physics and Technology, Institutsky lane 9, Dolgoprudny, Moscow region, 141700 Russia }
\affiliation{$^4$I. Physikalisches Institut, Justus-Liebig-Universit\"{a}t Gie\ss en, Heinrich-Buff-Ring 16, 35392 Gie\ss en, Germany}
\affiliation{$^5$Gagarin Research and Test Cosmonaut Training Center, 141160 Star City, Moscow Region, Russia}

\date{\today}
\begin{abstract}
Three-dimensional structure of complex (dusty) plasmas was investigated under long-term microgravity conditions in the~\mbox{International-Space-Station-based}~{Plasmakristall\mbox{-}4} facility.
The microparticle suspensions were confined in a polarity-switched dc discharge.
The experimental results were compared to the results of the molecular dynamics simulations with the interparticle interaction potential represented as a superposition of isotropic Yukawa and anisotropic quadrupole terms.
Both simulated and experimental data exhibited qualitatively similar structural features indicating the bulk liquid-like order with the inclusion of solid-like strings aligned with the axial electric field.
Individual strings were identified and their size spectrum was calculated.
The decay rate of the size spectrum was found to decrease with the enhancement of string-like structural features.
\end{abstract}


\maketitle
\newcommand{\pkfour}{PK\mbox{-}4}
\section{Introduction}
Complex plasmas \cite{fortov2005complex, morfill2009complex, FMBook, ILMRbook} are used in fundamental research as models for particle-resolved studies of generic classical condensed matter phenomena.
2D complex plasmas can be successfully studied under ground laboratory conditions \cite{Melzer2000, Nunomura2000, nosenko2009, nosenko2013}, whereas for the investigations of homogenous 3D systems, microgravity is required.
In the frame of a research program (\mbox{PKE-Nefedov} \cite{PKENefedov} and \mbox{PK-3 Plus} \cite{thomas2008complex} laboratories) on the International Space Station (ISS), such phenomena as freezing and melting transitions \cite{KlumovPU, KlumovPPCF2009, klumov2010EPL, khrapak2011, khrapak2012}, electrorheology \cite{ivlev2008electrorheology}, shock waves \cite{samsonov2003shock}, lane formation \cite{du2012EPLlane, du2012NJPlane}, etc. were investigated in 3D complex plasmas under long-term microgravity.
\\ \indent
Plasmakristall\mbox{-}4 (\pkfour)~\cite{pustylnik2016pk4} laboratory is a presently operational complex plasma facility on board the ISS.
It is intended for the microgravity investigations of anisotropic fluid phase of complex plasmas.
3D analysis of microparticle suspensions trapped by a polarity-switched discharge in the ground-based \pkfour~setup showed the presence of crystalline order \cite{Mitic08}.
In another laboratory experiment, the suspensions trapped in a \pkfour~chamber in a combined rf and dc discharge were investigated.
Their 3D analysis revealed coexistence of solid-, fluid- and string-like phases \cite{Mitic2013}.
\\ \indent
In this work, we report on the first comprehensive analysis of a 3D structure of a uniform string-fluid complex plasma.
Interest to the string-fluid state of complex plasmas arose after the discovery of their electrorheology \cite{ivlev2008electrorheology}.
``Conventional'' electro- \cite{Chen1992, Dassanayake2000} or magnetorheological fluids \cite{Ashour1996, deVicente2011} consist of suspensions of microparticles in nonconducting fluids.
They can drastically change their viscosity on the application of external electric or magnetic field. 
This property makes their use very promising in such applications as vibration control \cite{Stanway1996} or biomedicine \cite{Liu2001}.
The external field polarizes (magnetizes) the microparticles and induces additional dipole coupling between them.
Microparticles start to form strings in the direction of the induced dipoles.
In complex plasmas, electric field polarizes not the microparticles themselves, but rather their shielding ion clouds.
\\ \indent
Up to now, the observations of the isotropic-to-string-fluid transition in complex plasmas were either incomplete or to a certain extent controversial.
Electrorheology in plasmas was discovered in \cite{ivlev2008electrorheology}, where the structural analysis was performed using the anisotropic scaling indices.
Complementary analysis of the same experimental data with the help of bond-orientational-order parameters \cite{klumov2010EPL} showed that the external fast oscillating electric field caused crystallization of the microparticle suspension and not the transition from isotropic to string fluid.
In parabolic flight experiments of Ref.~\cite{ivlev2011electrorheology}, the microgravity conditions were achieved only for about $20$~s in each parabola.
Unlike \cite{ivlev2011electrorheology}, our work was done in an ISS-based facility.
This gave us sufficient time to investigate the 3D structure of our microparticle suspensions and to observe its evolution.
In the 3D structure of our suspension, we were able to identify the undoubtful signatures of strings.
Additionally, we supplemented our experimental data with molecular dynamics (MD) simulations.
Data from the experiments and simulations were processed using the same techniques.
\\ \indent
The paper is organized as follows: In section \ref{Sec:Exp}, we explain our experimental setup. 
In section \ref{Sec:Sim}, we describe the simulations we used to compare with our experiment.
In section \ref{Sec:Ana}, we explain the analysis techniques we employed both on experimental and simulated data.
In section \ref{Sec:RD}, we show and discuss the results of this data analysis and, in section \ref{Sec:Conc}, we draw the conclusions.
\\ \indent
\section{Experiment}
\label{Sec:Exp}
Experiments were performed in the flight model of \pkfour~setup on board the ISS \cite{pustylnik2016pk4}.
In \pkfour, a dc discharge plasma is produced inside a glass tube of $3$~cm diameter.
The discharge can be operated in continous dc as well as in polarity switching mode.
Typical polarity switching frequency of the order of hundreds Hz is still low enough for the dc discharge to stabilize within half of the switching period, but already high enough for the microparticles not to react on the polarity switching.
Maximal current in the dc mode as well as for one of the polarities in the polarity switching mode is 3 mA.
Gas pressure can be varied between $0.1$~and $2$~mbar.
\\ \indent
The working area of the tube has a length of about $20$~cm.
In this area, the microparticles can be illuminated by a laser sheet.
The scattered light is observed by the two video cameras.
\\ \indent
Melamine formaldehyde microparticles are injected into the plasma chamber outside of the working area and need to be transported before they can be seen by the cameras.
In this particular work, a dc trapping technique, in which the microparticles of $3.4$~$\mu$m diameter are transported by a dc discharge in argon (current $0.5$~mA, pressure $0.4$~mbar) and trapped by a polarity-switched discharge (current $0.5$~mA and frequency $500$~Hz), was used.
\\ \indent
\begin{figure}[t!]
\centering
\includegraphics[width=0.4\textwidth]{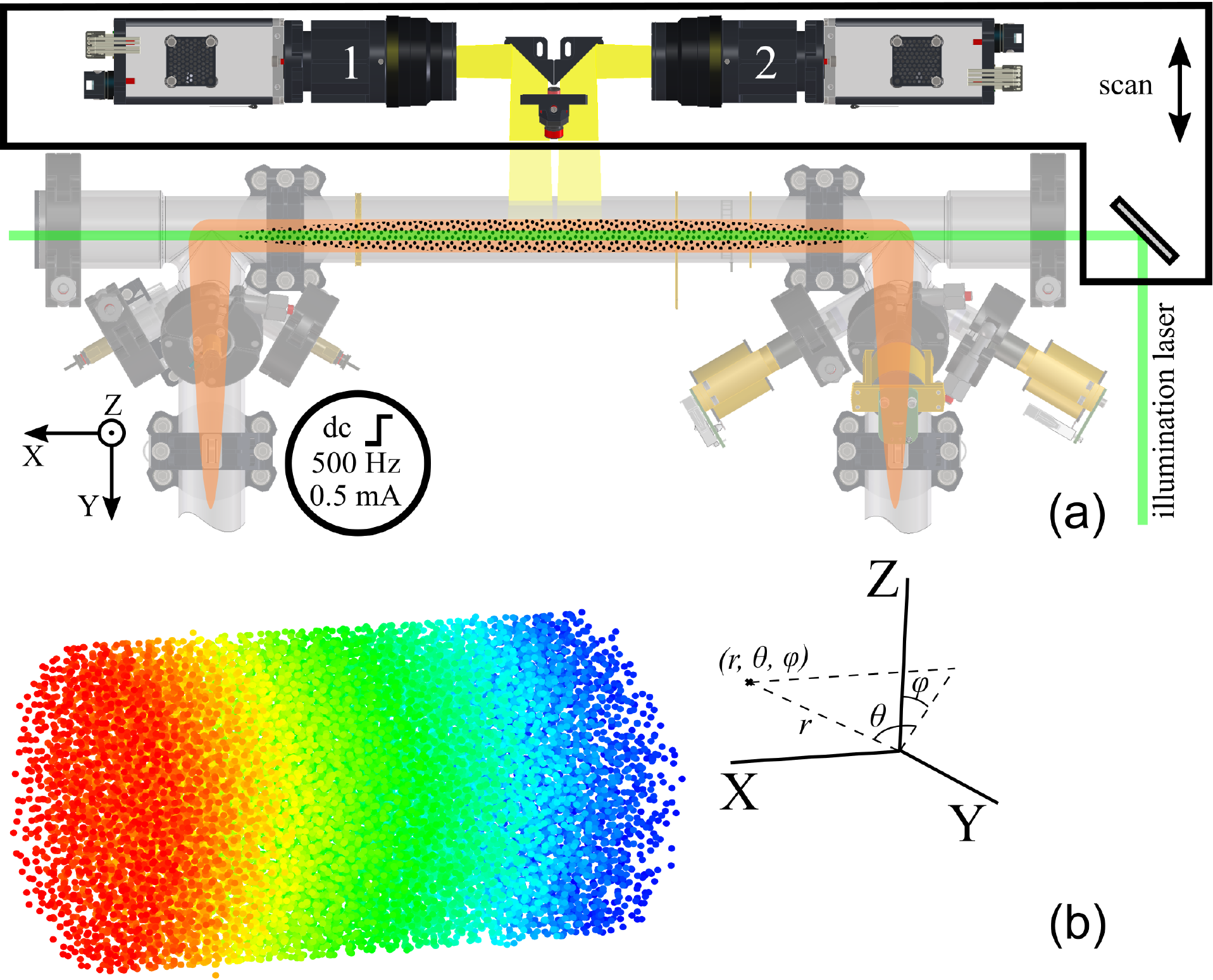}
\caption{
(a) Scheme of the experiment.
A large homogenous suspension of microparticles is trapped using the polarity-switched dc discharge.
The 3D structure of the suspension is reconstructed from a series of 2D slices obtained during the scan in the direction perpendicular to the image plane.
(b) An example of a 3D reconstruction of a microparticle suspension (color-coded by $X$~coordinate).
The suspensions exhibit cylindrical shape.}
\label{Fig: Scheme}
\end{figure}
The trapped microparticle suspension was scanned in the direction perpendicular to the image (Fig.~\ref{Fig: Scheme}a).
For this, the cameras and the laser optics were synchronously moved to keep the plane of the laser sheet always in the focus of the cameras.
The scanning speed of $v_{\rm scan}=0.9$~mm/s results in about $15~\mu$m separation between the consecutive frames along the Y axis.
Video data from camera~$1$ only were used in this work.
The field of view with the dimension $600\times 600$ pixels centered on the axis of the plasma chamber was used.
Axial position of this field of view was chosen in the rightmost part of the camera~$1$ image, where the illumination laser sheet has minimal thickness.
\\ \indent
Scanning allowed to reconstruct the 3D structure of the microparticle suspension.
This was done in the following way.
First, all the 2D images recorded during the scan were stacked into a 3D image with a voxel volume of $14.18\times 15.00\times 14.25~\mu$m$^3$.
In this 3D image, the objects consisting of voxels with the intensities above certain threshold were identified.
These objects correspond to the trajectories of individual microparticles while they are illuminated during the scan.
3D position of each microparticle was determined by weighting the intensities of the voxels belonging to the respective 3D object.
An example of a reconstructed 3D structure is shown in Fig.~\ref{Fig: Scheme}b.
Five scans with the interval of $\approx 2$~min between them were performed.
\\ \indent
We consider the data from the two scans: The first scan was performed right at the beginning of the experiment, whereas the second was performed about $6$~min later.
Diameter of the suspension decreased from $8.5$~mm to $7.2$~mm between those two scans, whereas the discharge current and pressure were kept constant.
The average microparticle number density was about $3.4\times10^4$~cm$^{-3}$.
The microparticle charge according to the measurements in \cite{Antonova2019} was about $2100$ elementary charges.
\\ \indent
\section{Simulations}
\label{Sec:Sim}
We compared the experimental observations with the results of MD simulations \cite{ivlev2008electrorheology, klumov2010EPL, ivlev2011electrorheology}.
In those simulations, the microparticle interaction potential was a superposition of Yukawa ($\phi_{\rm Y}$) and quadrupole-like ($\phi_{\rm Q}$) terms \cite{ivlev2008electrorheology}.
The terms are expressed as follows:
\begin{equation}
\phi_{\rm Y} = \frac{Q_{\rm d}^2}{4\pi\epsilon_0 d}\exp{\left(-\frac{d}{\lambda}\right)},
\end{equation}
where $Q_{\rm d}$ is the microparticle charge, $d$ is the interparticle distance, $\lambda$ is the Debye screening length, and
\begin{equation}
\phi_{\rm Q} = -0.43\frac{Q_{\rm d}^2}{4\pi\epsilon_0 d}\frac{M^2\lambda^2}{d^2}\left(3\cos^2\zeta-1\right),
\end{equation}
where $M$ is the thermal Mach number defined as a ratio of the drift velocity of ions to their thermal velocity and $\zeta$ is an angle between the external electric field and the vector connecting two microparticles.
The origin of the anisotropy is the fast oscillating electric field of a polarity-switched dc discharge.
The quadrupole-like term mimics the polarization of an ion cloud screening the microparticle.
The simulated system consisting of $45000$ particles was embedded in the Langevin thermostat like in \cite{KlumovPU, klumov2010EPL}.
\\ \indent
Each system was characterized by three dimensionless parameters: screening parameter $\kappa=D/{\lambda}$ ($D$ is the mean interparticle distance), Mach number $M$ and coupling parameter $\Gamma=\frac{Q_{\rm d}^2}{4\pi\epsilon_0 DT}$ ($T$ is the kinetic temperature of microparticles).
We used the following parameters: thermal ion Mach number $M=1.2$, screening parameter $\kappa=3.6$ and coupling parameter $\Gamma\simeq 10^3$.
In addition, we used the data for the Yukawa melt \cite{Khrapak2016} with $\kappa=3$ as a reference isotropic fluid.
Estimation based on the dust-free plasma density $2.1\times10^8$~cm$^{-3}$ for the experimental conditions under consideration yields $\kappa\approx 3.6$.
Ion Mach number for the electric field of $2.5$~V/cm estimated using the modified Frost formula \cite{khrapak2019frost} yields $M\approx 0.33$.
\\ \indent
\section{Data analysis}
\label{Sec:Ana}
Using 3D coordinates of microparticles (either obtained form the experiment or available from simulations), we analyzed the order in the respective systems.
Below, we list the quantities we used for the analysis.
\\ \indent
Pair corelation function was calculated in spherical coordinates as
\begin{widetext}
\begin{equation}
G(r,\theta,\phi)=\frac{1}{n_{\rm d}N}\sum^N_{\substack{i,j=1 \\ i \neq j}}\frac{\delta(r_{\rm ij}-r)\delta(\theta_{\rm ij}-\theta)\delta(\phi_{\rm ij}-\phi)}{4\pi r^2 \cos{\theta}},
\end{equation}
\end{widetext}
where $r$, $\theta$ and $\phi$ are the current spherical coordinates, $r_{\rm ij}$, $\theta_{\rm ij}$ and $\phi_{\rm ij}$ are the length, polar angle and azimuthal angle of a vector connecting microparticles $i$ and $j$, $N$ is the number of microparticles, $n_{\rm d}$ is the macroscopic microparticle number density.
About $6\times 10^3$~microparticles in the central part of the suspension were used for $G(r,\theta,\phi)$ calculation.
We have also defined several integrals of $G(r,\theta,\phi)$, namely a two-dimensional distribution 
\begin{equation}
G_{\rm \phi}(r,\theta)=\int^{2\pi}_0 G(r,\theta,\phi)d\phi, 
\end{equation} 
radial distribution function (RDF) 
\begin{equation}
g(r)=\int^{2\pi}_0\int^{\frac{\pi}{2}}_{-\frac{\pi}{2}} G(r,\theta,\phi)\cos{\theta}d\theta d\phi, 
\end{equation} 
cumulative RDF
\begin{equation}
N(<r)=4\pi n_{\rm d}\int_0^r{\xi^2g(\xi)d\xi}, 
\end{equation} 
and polar angle distribution function (PADF) 
\begin{equation}
p(\theta)=\int^{2\pi}_0\int^{r_1}_{0} 4\pi r^2 G(r,\theta,\phi)dr d\phi,
\end{equation} 
where $r_1$ is the radius of the first coordination sphere.
\\ \indent
The suspension exhibited a jerky drift in the axial direction. 
The axial velocity distribution of the microparticles measured between the scans can be characterized with the average $v_{\rm drift} = 0.2$~mm/s and full width at half maximum $\Delta v_{\rm fwhm}=0.7$~mm/s.
The drift weakens the correlations between the microparticles in the scanning direction.
We estimate the maximal radius $r_{\rm max}$, within which the correlations are not affected by the drift, as $r_{\rm max}/D=v_{\rm scan}/\sqrt{v_{\rm drift}^2+(\Delta v_{\rm fwhm}/2)^2}$.
For our experimental conditions with $D=0.32$~mm, $r_{\rm max}=0.79$~mm.
For the calculation of $G(r,\theta,\phi)$, we consider the maximal correlation radius equal to $1$~mm.
\\ \indent
Other characteristics we employed are bond angle distribution function (BADF) $P(\alpha)$, showing the probability of two nearest neighbors and central particle to form angle $\alpha$ \cite{Allen2017}, bond orientational order parameters (BOOP) (\cite{Steinhardt1981, Steinhardt1983, Mitu1982}) $q_4$ and $q_6$ and angular distribution of the nearest neighbours.
Since these are local characteristics, drift of the microparticle suspension should not strongly affect them.
\\ \indent
As we will show in section \ref{Sec:RD}, our microparticle suspensions exhibit the properties of string fluid and therefore contain the string-like clusters (SLCs).
To quantitatively characterize the SLCs, we identified them and calculated their size distributions.
The identification was performed in two steps:
First, we found the pairs of neighboring particles, whose bonds are almost parallel to the axis of the discharge tube.
The allowed angular deviation from the axis was estimated using the PADFs.
After that, the pairs were connected into SLCs with the help of \mbox{friend-of-friends} algorithm (see e.g. \cite{KlumovPU}).
Identification of the SLCs allowed us to calculate the RDF inside them.
\begin{figure}[t!]
\includegraphics[width=0.4\textwidth]{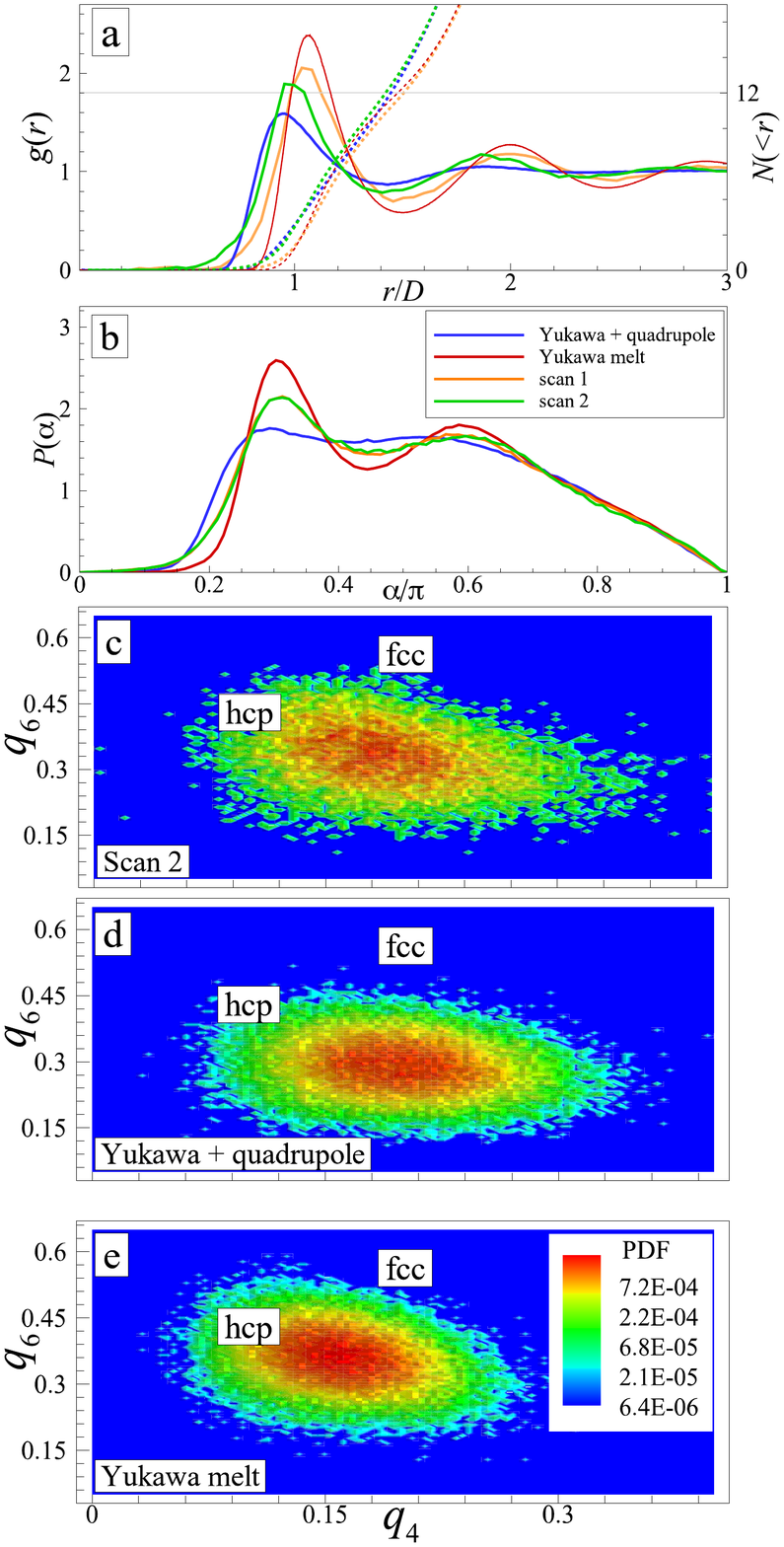}
\caption{One-dimensional correlation functions: (a) radial distribution function $g(r)$, (b) bond angle distribution function $P(\alpha)$ for two experimental data sets, simulated Yukawa melt and simulated anisotropic complex plasma and 2D probability distributions of microparticles on the $q_4\--q_6$ plane  for the experimental data (second scan) (c) and simulated systems: 
anisotropic complex plasma (d) and Yukawa melt (e). 
In panel (a), the cumulative RDFs $N(<r)$ are also plotted with dashed lines, revealing close packing (12 nearest neighbors in the first coordination shell) structures for the microparticle suspension.
$g(r)$, $P(\alpha)$ as well as the local orientational order from BOOP data clearly reveals liquid-like structure for all considered systems.}
\label{Fig:liquid}
\end{figure}
\begin{figure*}[t!]
\includegraphics[width=17.2cm]{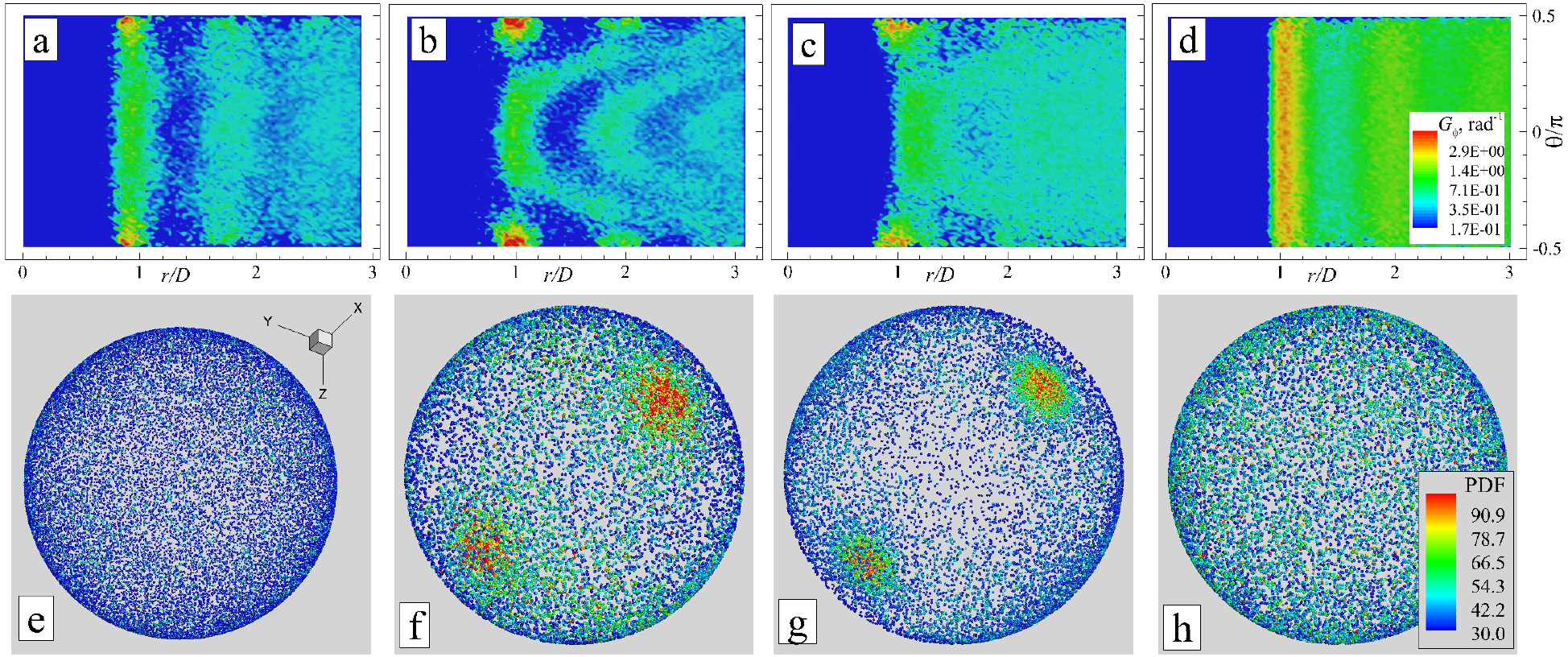}
\caption{Distribution $G_\phi$ and angular distributions of the nearest neighbors ($N_{\rm nn} = 12$) projected onto a sphere calculated for (a), (e) first experimental scan; (b), (f) second experimental scan; (c), (g) simulated anisotropic complex plasma and (d), (h) simulated Yukawa melt.
The peaks in $G_\phi$ at $\theta=\pm\frac{\pi}{2}$ in the experimental systems and for the simulated anisotropic plasma manifest the formation of axial strings.
For both experimental scans $D=0.32$~mm.
Angular distributions also demonstrate bond alignment along the $X$ axis for the second experimental system and simulated anisotropic system.
Yukawa melt is an isotropic system, as expected.}
\label{Fig: Gs}
\end{figure*}
\\ \indent
\section{Results and discussion}
\label{Sec:RD}
\begin{figure}
\includegraphics[width=0.45\textwidth]{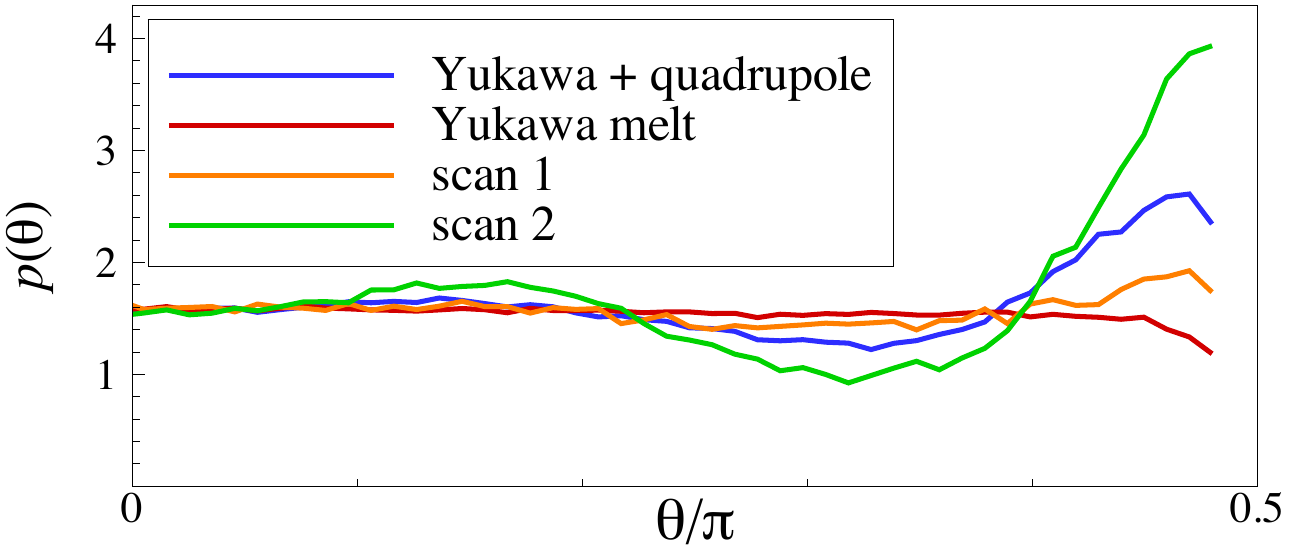}
\caption{Polar angle distribution function for all the four systems under consideration.
Peaks in the vicinity of  $\theta=\pm\frac{\pi}{2}$ reveal string-like features in the axial direction. 
The range of angles between $-\frac{\pi}{2}$ and $0$ is omitted due to symmetry.}
\label{Fig:padf}
\end{figure}

In Fig.~\ref{Fig:liquid}(a), $g(r)$ and the cumulative RDFs are plotted.
The latter function allows to estimate the mean nearest neighbours number $N_{\rm nn}$ in the first coordination shell.
Experimental $g(r)$ reveal nearly close packed structures ($N_{\rm nn} = 12$). 
Both correlation functions  $g(r)$ (Fig.~\ref{Fig:liquid}(b)) and $P(\alpha)$ (Fig.~\ref{Fig:liquid}(b)) exhibit liquid-like structures.
Analysis of our suspensions with the help of bond orientational order parameters (BOOP) (\cite{Steinhardt1981, Steinhardt1983, Mitu1982}) $q_4$ and $q_6$ confirms the liquid-like character of our suspensions (Fig.~\ref{Fig:liquid}(c)-(e)). 
Comparison of the first peaks of $g(r)$ with those for the Yukawa melt allows us to estimate the relative coupling parameter \cite{Khrapak2020} $\Gamma_{\rm m}/\Gamma \simeq 1.25$, where $\Gamma_{\rm m}$ is the coupling parameter at melting.
All quantities displayed in Fig.~\ref{Fig:liquid} exhibited liquid order in the experimental suspensions and even in the simulated anisotropic complex plasma.
However, the following, more detailed analysis shows clear indications of presence of the SLCs.  
\\ \indent
Figs.~\ref{Fig: Gs}(a) and~\ref{Fig: Gs}(b) show the $G_{\rm \phi}$ function for the two scans.
In both cases, correlations exhibit three peaks over $r/D$ corresponding to $1$, $2$ and $3$, respectively.
The polar angle correlations exhibit well defined peaks at $\theta=\pm\frac{\pi}{2}$ over three interparticle distances.
These peaks reveal strong correlations in the $X$ direction, suggesting string-like character of the suspension.
Fig.~\ref{Fig: Gs}(b) shows even more complicated arch-like structures at least for two interparticle distances, which may be interpreted as characteristics of interstring correlations.
Similar peaks at $\theta=\pm\frac{\pi}{2}$ and a very similar structure in $G_{\rm \phi}$ is observed in our MD-simulations (Fig.~\ref{Fig: Gs}(c)).
\\ \indent
Angular distribution of nearest neighbors exhibits similar behavior.
It is presented in Figs.~\ref{Fig: Gs}(e)-(h) in a projection onto a sphere.
It is almost isotropic for the first experimental scan and the Yukawa melt. 
For the second experimental scan and the simulated anisotropic complex plasma, the areas with enhanced density of nearest neighbors clearly concentrate in the polar regions.
\\ \indent
Fig.~\ref{Fig:padf} shows the PADFs for all the four systems under consideration.
Again, peaks in the vicinity of $\theta=\pm\frac{\pi}{2}$ are observed for the two experimental datasets and for the simulated anisotropic complex plasmas, whereas, Yukawa melt is isotropic.
We also note that RDF (Fig.~\ref{Fig:liquid}(a)) and BADF (Fig.~\ref{Fig:liquid}(b)) for the two experimental datasets look almost identical, whereas the PADF (Fig.~\ref{Fig:padf}), angular distribution of nearest neighbors and $G_{\rm \phi}(r,\theta)$ (Fig.~\ref{Fig: Gs}) show enhancement of the string-like features.
\\ \indent
\begin{figure}
\includegraphics[width=8.6cm]{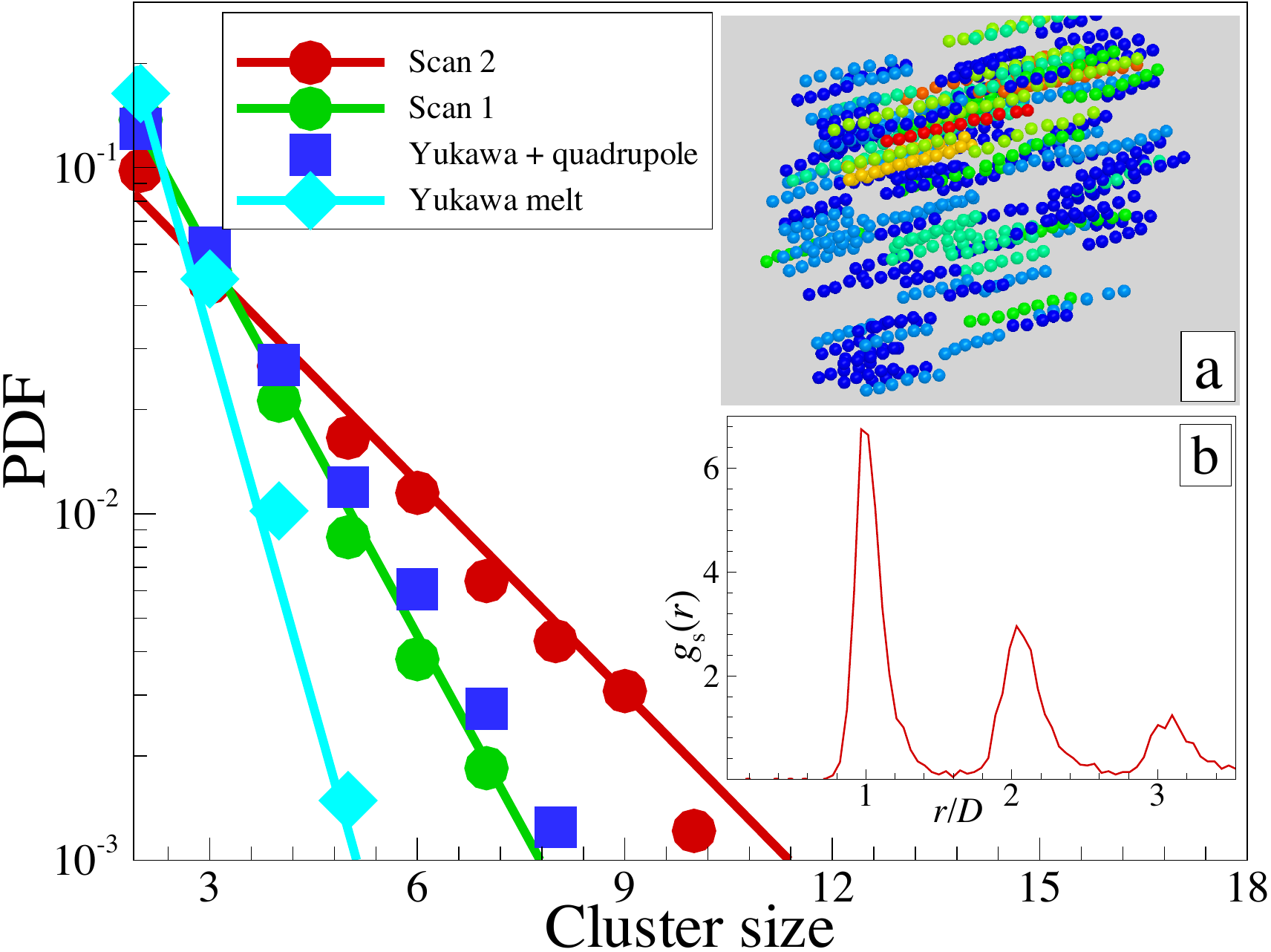}
\caption{
Probability distribution of string-like cluster (SLC) lengths for different systems, both experimental and simulated. 
All the spectra decay exponentially.
Anisotropic systems exhibit smaller decay rate compared to the isotropic Yukawa melt.
Inset (a) shows spatial distribution of the SLCs observed in experiment (second scan). 
The SLCs are color coded according to their size.
Inset (b) shows the radial distribution function $g_s(r)$ for the microparticles inside the strings.
Note that $g_s(r)$ exhibits solid-like order.}
\label{Fig: spectr}
\end{figure}
We were able to identify the individual SLCs (selected examples are shown in the inset (a) to Fig.~\ref{Fig: spectr}).
The allowed angular deviation of a bond direction from $X$ axis was estimated using the PADFs (Fig.~\ref{Fig:padf}) to be $\approx\frac{\pi}{10}$.
Fig.~\ref{Fig: spectr} shows the size spectra of the SLCs.
Size spectra of SLCs decay exponentially for all the systems under consideration, however, the decay rate in the anisotropic cases decreases significantly with respect to the isotropic Yukawa melt.
Also, the second scan, where the strings are much better developed, exhibits smaller decay rate than the first scan. 
\\ \indent
One-dimensional pair correlation function of the SLCs, $g_{\rm s}(r)$, for the second scan is shown in the inset (b) Fig.~\ref{Fig: spectr}.
The function is normalized to the total number of microparticles belonging to SLCs and their linear density in the strings.
We stress that this $g_s(r)$ exhibits crystalline order.
Therefore, the observed system represents bulk liquid with the inclusion of solid-like strings aligned along the discharge tube axis.
\\ \indent
We are at the moment unable to explain the observed enhancement of the string-like order with time in our suspension.
There can be various physical mechanisms responsible for it.
Experiments in the PK-3 Plus microgravity laboratory showed that the crystallization of microparticle suspension can last several minutes \cite{pk3plusTeam}.
Also, we cannot exclude at the moment that the observed evolution is not the consequence of spatial non-uniformity of the suspension:
Its better ordered parts could arrive into the field of view of the cameras at later times due to the drift.
Another mechanism could be associated with decrease of the suspension diameter in time.
It has been shown earlier in \cite{Zobnin2018} that the microparticle suspensions immersed in the dc discharge can affect the background plasma.
This can in principle enhance the anisotropic attractive term of the microparticle interaction potential.
Understanding the physical mechanisms responsible for the variations of the string-like order in our suspensions will be the subject of future investigations.
\\ \indent
\section{Conclusions}
\label{Sec:Conc}
To summarize, we have performed comprehensive 3D structural analysis of the microparticle suspensions in the dc plasma with fast oscillating electric field under microgravity conditions.
Although radial distribution, bond angle distribution functions and rotational invariants $q_4$ and $q_6$ exhibited fluid order, angular functions derived from the 3D pair correlation function showed clear axial peaks.
We were able to identify individual strings and measure their size spectrum.
Decay rate of the size spectrum of the string-like clusters was found to decrease on the increase of the axial peak.
Radial distribution function measured inside the strings exhibited solid-like order.
Similar features were shown to appear in the MD simulations on the addition of the quadrupole-like component to the microparticle interaction potential.
Finally, we observed for the first time the 3D structure of a string-fluid complex plasma.
This opens up the possibilities for novel experiments, e.g., observations of isotropic-to-string-fluid transition, which will be explored using the \pkfour~apparatus.
\\ \indent
The authors thank Dr. C. Knapek for careful reading of the manuscript.
All authors greatly acknowledge the joint ESA-Roscosmos ``Experiment Plasmakristall-4'' on-board the International Space Station.
This work was supported in part by DLR/BMWi grants 50WM1441 and 50WM1442.
B. K. was supported by the German Academic Exchange Service (DAAD) with funds from the German Aerospace Center (DLR).
%

\end{document}